\documentclass[fleqn,usenatbib]{mnras}

\usepackage{newtxtext,newtxmath}
\usepackage[T1]{fontenc}
\usepackage{graphicx}
\usepackage{amsmath}
\usepackage{bm}
\usepackage{xcolor}
\usepackage{hyperref}
\usepackage{booktabs}
\usepackage{multirow}
\graphicspath{{figures/}}

\newcommand{\tref}{T_{\rm ref}}
\newcommand{\tns}{T_{\rm NS}}

\newcommand{\tcmb}{T_{\rm CMB}}
\newcommand{\tfg}{T_{\rm FG}}
\newcommand{\nuref}{\nu_{\rm ref}}
\newcommand{\reach}{\textsc{reach}}
\newcommand{\edges}{\textsc{edges}}

\newcommand{\hera}{\textsc{hera}}
\newcommand{\ska}{\textsc{ska}}
\newcommand{\lofar}{\textsc{lofar}}
\newcommand{\mwa}{\textsc{mwa}}
\newcommand{\jax}{\textsc{jax}}
\newcommand{\blackjax}{\textsc{BlackJAX}}

\newcommand{\aoflagger}{\textsc{AOFlagger}}
\newcommand{\fgivenx}{\textsc{fgivenx}}
\newcommand{\treftk}{T_{{\rm ref},t}^{(k)}}

\title{Scalable Bayesian data curation for next-generation radio experiments}

\author[S.\,A.\,K. Leeney et al.]{
S.\,A.\,K. Leeney,$^{1,2}$\thanks{E-mail: sakl2@cam.ac.uk}
E. de Lera Acedo,$^{1,2}$
W.\,J. Handley,$^{2,3}$
H.\,T.\,J. Bevins,$^{1}$
G. Allen,$^{4}$
\newauthor
D. Anstey,$^{1,2}$
K. Artuc,$^{1,2}$
G. Bernardi,$^{7}$
M. Bucher,$^{4,6}$
S. Carey,$^{1}$
\newauthor
J. Cavillot,$^{8}$
R. Chiello,$^{9}$
A.\,S. Chu,$^{1,2}$
W. Croukamp,$^{4}$
J. Cumner,$^{1,2}$
\newauthor
S. Dasgupta,$^{2,3}$
A.\,K. Dash,$^{1,2}$
D.\,I.\,L. de Villiers,$^{4}$
J. Dhandha,$^{2,3}$
A. Dragovic,$^{1,2}$
\newauthor
J.\,A. Ely,$^{1}$
A. Fialkov,$^{2,3}$
T. Gessey-Jones,$^{1,2}$
C. Kirkham,$^{1,2}$
G. Kulkarni,$^{10}$
\newauthor
A. Magro,$^{5}$
P.\,D. Meerburg,$^{11}$
S. Mittal,$^{1,2}$
D. Molnar,$^{1,2}$
R.\,S. Patel,$^{1,2}$
\newauthor
J.\,H.\,N. Pattison,$^{1,2}$
S. Pegwal,$^{4}$
C.\,M. Pieterse,$^{4}$
J.\,R. Pritchard,$^{12}$
G.\,M.\,Z. Rajpoot,$^{1,2}$
\newauthor
N. Razavi-Ghods,$^{1}$
D. Robins,$^{1,2}$
I.\,L.\,V. Roque,$^{16}$
A. Saxena,$^{1,2}$
K.\,H. Scheutwinkel,$^{1,2}$
\newauthor
E. Shen,$^{1,2}$
P.\,H. Sims,$^{1,2}$
M. Spinelli,$^{13,14}$
J.\,L. Tutt,$^{1,2}$
J. Zhu$^{1,15}$
\\
$^{1}$Astrophysics Group, Cavendish Laboratory, University of Cambridge, J.\,J. Thomson Avenue, Cambridge, CB3 0US, UK\\
$^{2}$Kavli Institute for Cosmology in Cambridge, University of Cambridge, Madingley Road, Cambridge, CB3 0HA, UK\\
$^{3}$Institute of Astronomy, University of Cambridge, Madingley Road, Cambridge, CB3 0HA, UK\\
$^{4}$Department of Electrical and Electronic Engineering, Stellenbosch University, Stellenbosch, 7602, South Africa\\
$^{5}$Institute of Space Sciences and Astronomy, University of Malta, Msida, Malta, MSD 2080, Malta\\
$^{6}$Laboratoire AstroParticule et Cosmologie, Universit\'e Paris-Cit\'e, 10 Rue Alice Domon et L\'eonie Duquet, Paris, 75013, France\\
$^{7}$INAF-Istituto di Radio Astronomia, Via Gobetti 101, Bologna, 40129, Italy\\
$^{8}$Antenna Group, Universit\'e catholique de Louvain, Louvain-la-Neuve, 1348, Belgium\\
$^{9}$Physics Department, University of Oxford, Parks Road, Oxford, OX1 3PU, UK\\
$^{10}$Department of Theoretical Physics, Tata Institute of Fundamental Research, Homi Bhabha Road, Mumbai, 400005, India\\
$^{11}$Faculty of Science and Engineering, University of Groningen, Nijenborgh 4, Groningen, 9747 AG, Netherlands\\
$^{12}$Max Planck Institute for Radio Astronomy, Auf dem H\"ugel 69, 53121 Bonn, Germany\\
$^{13}$Observatoire de la C\^ote d'Azur, Nice, France\\
$^{14}$Department of Physics and Astronomy, University of the Western Cape, Robert Sobukhwe Road, Bellville, 7535, South Africa\\
$^{15}$National Astronomical Observatory, Chinese Academy of Sciences, Beijing, 100101, China\\
$^{16}$SLAC National Accelerator Laboratory, Stanford University, Menlo Park, California 94025-7015, USA
}

\date{Accepted XXX. Received YYY; in original form ZZZ}

\pubyear{2026}

\begin{document}
\label{firstpage}
\pagerange{\pageref{firstpage}--\pageref{lastpage}}
\maketitle

\begin{abstract}
Next-generation radio telescopes produce data volumes that preclude manual quality assessment, yet data curation remains essential for science. We present a general, fully automatic Bayesian anomaly-detection method for radio science experiments in which data curation is performed inside the inference: a latent anomaly indicator is marginalised in the likelihood rather than converted into an external pre-flag. Implemented in JAX with GPU-accelerated inference, the pipeline assigns probabilistic data-curation scores without prior knowledge and requires no thresholds, manual inspection, or subjective decisions. We demonstrate the method on the Radio Experiment for the Analysis of Cosmic Hydrogen (REACH), applying it to 4655 observations (one year of REACH data). The pipeline assigns scores across time and frequency, enabling identification of the optimal observations to carry forward into scientific inference while reducing the risk that contaminated data bias the result. In doing so, it simultaneously recovers weather-driven systematics, instrument-component drifts, and narrow-band radio-frequency interference, while revealing complex dependencies between data quality and environmental or instrumental state that would be difficult to uncover by manual curation alone. This turns data curation from an external manual bottleneck into autonomous, inference-level infrastructure for the Square Kilometre Array era.
\end{abstract}


% ============================================================================
\section{Introduction}
\label{sec:introduction}
% ============================================================================

Modern observational astronomy has entered a regime in which data acquisition outstrips any realistic capacity for human inspection or complete retention. The first pulsar was detected in 1967 by visual inspection of pen-chart recordings \citep{Hewish1968}. Instrumentation and analysis have advanced enormously since then, yet human judgement remains embedded in most data-quality workflows. Contemporary facilities now operate at data rates that make such oversight impossible. The \ska\ is expected to generate more than an exabyte of raw data per day \citep{Dewdney2009,Bonaldi2021}; \lofar\ has already accumulated petabyte-scale interferometric datasets \citep{vanHaarlem2013,Shimwell2019,Norris2011}; and experiments such as the \mwa\ \citep{Wayth2018} and \hera\ \citep{DeBoer2017,LaPlante2021} routinely record terabytes per night. Gravitational-wave observatories already generate terabytes of strain data per observing run and face growing data-characterisation demands as detector networks expand \citep{Prathaban2025,Yallup2025gw}. Space-based missions such as the James Webb Space Telescope generate comparably demanding data products \citep{Gardner2006,Whitler2025}. At such scales, substantial fractions of the raw stream are discarded before scientific analysis simply because storage and bandwidth cannot support full retention. The data that remain must still be curated, assessed for quality, screened for contamination, and selected for inference, before any scientific conclusions can be drawn. Data curation is therefore no longer a peripheral bookkeeping task; it is a scientific bottleneck that directly limits the return on observational investment.

The consequences of poor curation are severe in both directions: contaminated data that are retained can bias parameter inference, while aggressive rejection of suspect data discards genuine information and introduces selection effects \citep{Leeney2025snia}. In radio astronomy, the standard automated response has been interference flagging outside the scientific model. Algorithms such as \aoflagger\ \citep{Offringa2010,Offringa2012}, \textsc{ssins} \citep{Wilensky2019}, and related heuristic and deep-learning classifiers \citep{Fridman2001,Akeret2017,Kerrigan2019} are highly effective at identifying radio-frequency interference in interferometric visibilities, but they are designed to excise interference in individual time--frequency samples rather than to assign statistically interpretable quality scores to whole nights, observing seasons, or instrumental configurations. Because these decisions are made outside the likelihood, the uncertainty associated with curation cannot be propagated into the final posterior. Even in workflows with automated diagnostics, analysts typically still set thresholds, inspect plots, and make subjective decisions about what to retain. Such human-in-the-loop curation is increasingly difficult to reproduce, difficult to scale, and difficult to defend statistically \citep{Sims2025}.

A different approach is to treat curation as part of the inference problem itself. The Bayesian anomaly-detection formalism introduced by \citet{Leeney2023} embeds contamination directly in the likelihood through a mixture of a physically motivated clean branch and a broad anomaly branch. Each datum is assigned a posterior anomaly probability, and uncertainty about contamination is marginalised rather than converted into a hard mask. This statistical principle is the central contribution of the present work. The formalism has already been extended to improve sensitivity to transient contamination \citep{Anstey2024rfi}, and the same anomaly-detection principle has been generalised beyond radio astronomy to the automated curation of Type Ia supernova light curves for cosmological inference \citep{Leeney2025snia}. Bayesian anomaly detection thus offers a domain-general route to automated, uncertainty-aware data curation.

We use the Radio Experiment for the Analysis of Cosmic Hydrogen (\reach) as a demanding proving ground. \reach\ is a global 21-cm experiment targeting the sky-averaged signal from the cosmic dawn and epoch of reionisation \citep{deLeraAcedo2022,Furlanetto2006,Pritchard2012,Mesinger2019}; the broader astrophysical context of the signal is reviewed elsewhere \citep{Barkana2016,Cohen2017,Fialkov2014,GesseyJones2022,GesseyJones2025,Pochinda2024,Barkana2018,Liu2019,Mittal2022,Brandenberger2019,Dhandha2025}. The measurement requires extracting a signal at the level of tens to hundreds of millikelvin from foregrounds four to five orders of magnitude brighter \citep{Shaver1999,deOliveiraCosta2008}, and the debate around the claimed \edges\ detection underscored how strongly conclusions depend on robust treatment of systematics \citep{Bowman2018,Hills2018,Singh2022}. The \reach\ collaboration has developed an extensive Bayesian analysis stack spanning forward modelling, calibration, anomaly-aware inference, and systematic treatments \citep{Leeney2023,Anstey2021,Anstey2022inform,Anstey2023,Bevins2021maxsmooth,Bevins2021globalemu,Pagano2023,Pattison2023,Shen2021,Shen2022ion,Shen2024flex,Saxena2023,Saxena2024,Mittal2024,Mittal2026,Carter2025,Pattison2025env,Pattison2026,Roque2021,Roque2025,Kirkham2024,Kirkham2025,Kirkham2026,Allen2026,Cumner2024,GesseyJones2024,Sims2025,Sims2025b,Leeney2025ml}, supported by mature inference and visualisation infrastructure \citep{Handley2015polychord,Handley2018fgivenx,Handley2019anesthetic,Scheutwinkel2022a,Scheutwinkel2023,Tutt2026,Leeney2025jaxbandflux}. The experiment is thus the application studied here, not the limit of the method's intended scope.

In this work we apply Bayesian anomaly detection to 4655 \reach\ observations and demonstrate fully automatic, uncertainty-aware data curation without thresholds, manual inspection, or subjective decisions. The emphasis is on moving curation from external pre-flagging into the likelihood, so that contaminated observations are marginalised over during inference rather than removed by a separate decision. This framing is essential for unbiased science analyses at next-generation data volumes, where manual choices cannot scale and hard thresholds are difficult to defend statistically.

The structure of this paper is as follows. Section~\ref{sec:theory} describes the Bayesian anomaly-detection formalism and the model used for the \reach\ application. Section~\ref{sec:method} details the implementation. Section~\ref{sec:results} presents the results. Section~\ref{sec:discussion} discusses the implications for radio astronomy and for automated curation more generally. Section~\ref{sec:conclusion} summarises our conclusions.

% ============================================================================
\section{Theory}
\label{sec:theory}
% ============================================================================

This section presents the statistical framework for Bayesian anomaly detection applied to data curation, following the formalism of \citet{Leeney2023}. We first summarise the Bayesian inference framework (\S\,\ref{subsec:bayesian_overview}), then derive the anomaly-detection likelihood in general terms (\S\,\ref{subsec:anomaly_formalism}), and finally describe the physical models to which it is applied in this work: the \reach\ foreground model (\S\,\ref{subsec:foreground_model}), the compression to a reference-frequency summary (\S\,\ref{subsec:compression}), and the smooth population model for the compressed quantity (\S\,\ref{subsec:population_model}).

% ----------------------------------------------------------------------------
\subsection{Bayesian inference and nested sampling}
\label{subsec:bayesian_overview}
% ----------------------------------------------------------------------------

We work within the standard Bayesian framework. For model parameters $\boldsymbol{\theta}$ and data $\mathcal{D}$, the posterior distribution under model $\mathcal{M}$ is
\begin{equation}
\label{eq:bayes}
\mathcal{P}(\boldsymbol{\theta} \mid \mathcal{D}, \mathcal{M}) = \frac{\mathcal{P}(\mathcal{D} \mid \boldsymbol{\theta}, \mathcal{M})\, \pi(\boldsymbol{\theta} \mid \mathcal{M})}{\mathcal{Z}_{\mathcal{M}}},
\end{equation}
where $\mathcal{P}(\mathcal{D} \mid \boldsymbol{\theta}, \mathcal{M})$ is the likelihood, $\pi(\boldsymbol{\theta} \mid \mathcal{M})$ is the prior, and
\begin{equation}
\label{eq:evidence}
\mathcal{Z}_{\mathcal{M}} \equiv \mathcal{P}(\mathcal{D} \mid \mathcal{M}) = \int \mathcal{P}(\mathcal{D} \mid \boldsymbol{\theta}, \mathcal{M})\, \pi(\boldsymbol{\theta} \mid \mathcal{M})\, \mathrm{d}\boldsymbol{\theta}
\end{equation}
is the Bayesian evidence \citep{MacKay2003,Jeffreys1961}. The evidence is central to model comparison: the ratio $\mathcal{Z}_{\mathcal{M}_1} / \mathcal{Z}_{\mathcal{M}_2}$ constitutes a Bayes factor encoding the relative support for two competing models \citep{Kass1995,Trotta2008}.

All numerical inference in this work is performed using GPU-accelerated nested sampling \citep{Skilling2006}, specifically the nested slice sampling algorithm implemented in \blackjax\ \citep{Cabezas2024,Yallup2026nss}. Nested sampling simultaneously delivers posterior samples and an estimate of $\mathcal{Z}_{\mathcal{M}}$, making it well suited to problems in which model comparison is as important as parameter estimation \citep{Handley2015polychord,Feroz2009}. For reviews of nested sampling in the physical sciences, see \citet{Skilling2006} and \citet{Ashton2022}.

% ----------------------------------------------------------------------------
\subsection{Bayesian anomaly detection}
\label{subsec:anomaly_formalism}
% ----------------------------------------------------------------------------

The core idea, introduced by \citet{Leeney2023}, is to embed data curation directly in the likelihood rather than treating it as an external preprocessing step. Consider a dataset $\mathcal{D} = \{\mathcal{D}_i\}_{i=1}^N$ modelled by a physical forward model with parameters $\boldsymbol{\theta}$ and per-datum likelihood $\mathcal{L}_i(\boldsymbol{\theta})$. For each datum we introduce a latent binary indicator $\epsilon_i$:
\begin{itemize}
\item $\epsilon_i = 1$: the datum is explained by the physical model;
\item $\epsilon_i = 0$: the datum is treated as contaminated.
\end{itemize}
The per-datum piecewise likelihood is
\begin{equation}
\label{eq:anomaly_piecewise}
\mathcal{P}(\mathcal{D}_i \mid \boldsymbol{\theta}, \epsilon_i) =
\begin{cases}
\mathcal{L}_i(\boldsymbol{\theta}), & \epsilon_i = 1, \\[4pt]
\Delta^{-1}\,\mathbf{1}(0 < \mathcal{D}_i < \Delta), & \epsilon_i = 0,
\end{cases}
\end{equation}
where $\mathbf{1}(\cdot)$ denotes the indicator function and $\Delta$ is the support width of a uniform anomaly distribution, set to approximately the maximum range of the data: $\epsilon_i = 1$ labels the clean branch, while $\epsilon_i = 0$ labels the anomalous branch, represented by a broad uniform likelihood. Contaminated data are thus assigned a broad, uninformative likelihood that does not constrain the model parameters.

The prior on the corruption status is Bernoulli:
\begin{equation}
\label{eq:bernoulli_prior}
\mathcal{P}(\epsilon_i) = p_a^{\,1-\epsilon_i}\,(1-p_a)^{\,\epsilon_i},
\end{equation}
where $p_a$ is the prior probability that any given datum is anomalous. Combining the piecewise likelihood with this prior and marginalising analytically over the latent indicator yields the per-datum marginal likelihood
\begin{equation}
\label{eq:channel_marginal}
\mathcal{P}(\mathcal{D}_i \mid \boldsymbol{\theta}, p_a) = (1-p_a)\,\mathcal{L}_i(\boldsymbol{\theta}) + p_a\,\Delta^{-1}\,\mathbf{1}(0 < \mathcal{D}_i < \Delta).
\end{equation}
The full dataset likelihood is the product over data:
\begin{equation}
\label{eq:obs_likelihood}
\mathcal{P}(\mathcal{D} \mid \boldsymbol{\theta}, p_a) = \prod_i \mathcal{P}(\mathcal{D}_i \mid \boldsymbol{\theta}, p_a).
\end{equation}
Contaminated data are downweighted at the likelihood level rather than excised by a separate algorithm. The posterior responsibility of the anomaly branch for each datum,
\begin{equation}
\label{eq:channel_responsibility}
\mathcal{P}(\epsilon_i = 0 \mid \mathcal{D}_i, \boldsymbol{\theta}) = \frac{p_a\,\Delta^{-1}}{(1-p_a)\,\mathcal{L}_i + p_a\,\Delta^{-1}},
\end{equation}
is a natural output of the inference and serves as the anomaly score for datum $i$. This is the central quantity for data curation: each datum receives a posterior probability of being contaminated, and this uncertainty propagates into all downstream inference without human intervention.

% ----------------------------------------------------------------------------
\subsection{The REACH foreground model}
\label{subsec:foreground_model}
% ----------------------------------------------------------------------------

The full \reach\ forward model expresses the foreground contribution to the antenna temperature as a sum over regional amplitude and spectral-index masks convolved with a precomputed chromatic response tensor \citep{Anstey2021,Anstey2023,Tutt2026}. In the notation of \citet{Tutt2026},
\begin{equation}
\label{eq:full_foreground}
\tfg(\nu, t) = \sum_{i=1}^{N_\alpha}\sum_{j=1}^{N_\beta} \alpha_i\, K_{ij}(\nu, t)\, \left(\frac{\nu}{\nu_0}\right)^{-\beta_j} + \tcmb,
\end{equation}
where $K_{ij}(\nu,t)$ is the chromatic response tensor, $\alpha_i$ are amplitude parameters, and $\beta_j$ are spectral-index parameters.

In this work we adopt the simplest non-trivial case ($N_\alpha = N_\beta = 1$). After subtraction of $\tcmb$, the per-observation foreground model is
\begin{equation}
\label{eq:one_order}
m_t(\nu;\, A_t, \beta_t) = A_t\, K_t(\nu)\, \left(\frac{\nu}{\nu_0}\right)^{-\beta_t},
\end{equation}
where $A_t$ is a per-observation amplitude, $\beta_t$ is a spectral index, and $K_t(\nu)$ is the observation-specific chromaticity function. The per-channel clean likelihood for observation $t$ at frequency $\nu_f$ is then
\begin{equation}
\label{eq:clean_likelihood}
\mathcal{L}_{tf}(\boldsymbol{\theta}_t) = \frac{1}{\sqrt{2\pi\,\sigma_t^2}} \operatorname{exp}\!\left[-\frac{(\mathcal{D}_{tf} - m_t(\nu_f;\, A_t, \beta_t))^2}{2\,\sigma_t^2}\right],
\end{equation}
where $\sigma_t$ is a per-observation noise scale inferred jointly with the foreground parameters. The Bayesian anomaly-detection framework of \S\,\ref{subsec:anomaly_formalism} is applied to this likelihood, with the data index $i$ running over frequency channels $f$ within each observation.

The next two subsections define the two-step compression used to make population-level curation tractable. The foreground model above is required because anomaly scores must be measured relative to a clean spectral prediction, but the quantities needed for selecting whole observations are lower dimensional than the full frequency spectrum. We therefore compress each per-observation posterior to a reference-frequency foreground temperature, carrying posterior draws rather than only a best-fitting value so that uncertainty from the spectral fit is retained. These draws then form the inputs to a second, population-level model of the expected $\tref$ variation with LST, against which whole observations can be assigned anomaly probabilities.

% ----------------------------------------------------------------------------
\subsection{Compression to a reference-frequency statistic}
\label{subsec:compression}
% ----------------------------------------------------------------------------

Rather than carrying the full spectrum forward, we compress each observation to a posterior over a single physically meaningful quantity, $\tref$, defined at a fixed reference frequency $\nuref = 75$\,MHz. This value is chosen because it lies near the middle of the $50$--$130$\,MHz observing band, away from the band edges. For every posterior draw $k$, the foreground model is re-evaluated at $\nuref$:
\begin{equation}
\label{eq:tref_draw}
\treftk = m_t(\nuref;\, A_t^{(k)}, \beta_t^{(k)}).
\end{equation}
Defining the chromaticity ratio $R_t(\nu) \equiv K_t(\nu) / K_t(\nuref)$, the foreground model can be rewritten as
\begin{equation}
\label{eq:model_tref}
m_t(\nu;\, \tref, \beta_t) = \tref\, R_t(\nu)\, \left(\frac{\nu}{\nuref}\right)^{-\beta_t}.
\end{equation}
For each observation we propagate $K = 256$ posterior draws of $\tref$ together with summary statistics.

% ----------------------------------------------------------------------------
\subsection{Population-level model for \texorpdfstring{$\tref(\mathrm{LST})$}{Tref(LST)}}
\label{subsec:population_model}
% ----------------------------------------------------------------------------

The ensemble of compressed $\tref$ posteriors is modelled as a smooth periodic function of local sidereal time using a Fourier basis. For harmonic order $H$,
\begin{equation}
\label{eq:fourier}
\mu_t \equiv \tref^{\rm smooth}(l_t) = c_0 + \sum_{h=1}^{H} \left[a_h \cos\!\left(\frac{2\pi h\, l_t}{24}\right) + b_h \sin\!\left(\frac{2\pi h\, l_t}{24}\right)\right],
\end{equation}
where $l_t$ is the LST of observation $t$ in hours. Bayesian anomaly detection (\S\,\ref{subsec:anomaly_formalism}) is applied again at the observation level: each propagated $\tref$ draw is modelled as a mixture of a Gaussian branch centred on $\mu_t$ and a broad contamination component. The per-draw mixture likelihood is
\begin{multline}
\label{eq:per_draw_mixture}
\mathcal{P}\!\left(\treftk \mid \mu_t, p_a, \sigma_{\rm int}\right) = (1-p_a)\, \mathcal{N}\!\left(\treftk \mid \mu_t,\; \sigma_{t,\mathrm{ker}}^2 + \sigma_{\rm int}^2\right) \\
+ p_a\, \Delta^{-1},
\end{multline}
where $\sigma_{\rm int}$ is a global intrinsic-scatter parameter and $\sigma_{t,\mathrm{ker}}$ captures the width of the compressed posterior for observation $t$. Averaging over the $K$ propagated draws gives the observation-level likelihood
\begin{equation}
\label{eq:draw_marginalisation}
p_t \approx \frac{1}{K} \sum_{k=1}^{K} \mathcal{P}\!\left(\treftk \mid \mu_t, p_a, \sigma_{\rm int}\right),
\end{equation}
and the full population likelihood is $\mathcal{P}(\mathcal{D} \mid \boldsymbol{c}, p_a, \sigma_{\rm int}) = \prod_t p_t$, where $\boldsymbol{c} = \{c_0, a_h, b_h\}$.

The Fourier order $H$ is not fixed a priori. We compare models $\mathcal{M}_H$ for $H = 1, \dots, 6$ using the Bayesian evidence $\mathcal{Z}_H$. The preferred model is selected automatically by the evidence, with the Bayesian Occam penalty \citep{MacKay2003,Jeffreys1961} guarding against overfitting.

% ============================================================================
\section{Method}
\label{sec:method}
% ============================================================================

This section describes the practical implementation of the two-stage pipeline, the dataset to which it is applied, and the computational configuration used for the analysis presented in this paper.

% ----------------------------------------------------------------------------
\subsection{The REACH first-year dataset}
\label{subsec:dataset}
% ----------------------------------------------------------------------------

The pipeline is applied to the \reach\ first-year observing dataset, comprising observations recorded at the Karoo Radio Astronomy Reserve in South Africa \citep{deLeraAcedo2022}. \reach\ is a total-power drift-scan radiometer: each observation is an antenna-temperature spectrum of the radio sky, formed by integrating the sky brightness over the visible hemisphere through the antenna beam as the Earth rotates. Across the $50$--$130$\,MHz band used here, these spectra are dominated by diffuse Galactic synchrotron emission, which is intrinsically smooth and approximately power-law in frequency, with spatially varying amplitude and spectral index \citep{Shaver1999,deOliveiraCosta2008}. The beam response is also frequency dependent, so different angular regions of this anisotropic sky contribute with different weights at different frequencies. We refer to this coupling between the frequency-dependent beam and the structured sky as chromaticity; it imprints additional spectral structure on the otherwise smooth synchrotron foreground and is encoded in the chromaticity function used below.

After initial data selection, the dataset contains $T = 4655$ individual drift-scan spectra spanning a frequency band of $50$--$130$\,MHz, distributed across the months of March through December and corresponding to one year of REACH observing. Each observation corresponds to a single spectrum acquired at a specific LST and calibrated through the \reach\ noise-wave framework \citep{Roque2025,Roque2021}.

Given the size of this dataset, a fully joint fit over all observations simultaneously would introduce several thousand observation-level nuisance parameters, rendering direct sampling computationally prohibitive. We therefore adopt a two-stage approximation in which each observation is first analysed independently (stage~1), compressed to a summary statistic, and then modelled collectively (stage~2). The stage-1 anomaly downweighting is carried into stage~2 indirectly through the compressed posteriors, so the pipeline is not anomaly-blind at the population level despite the factorisation. We regard this as a reasonable compromise between statistical fidelity and computational tractability for the present dataset.

The chromaticity function $K_t(\nu)$ for each observation is precomputed from beam-convolved sky integrals following the \reach\ forward-modelling framework \citep{Anstey2021,Anstey2023,Tutt2026}. Weather metadata, including daily precipitation and temperature, are obtained from the Open-Meteo reanalysis archive for the Karoo site and joined to the observation frame by calendar date. Calibration metadata, including the noise-source temperature $\tns$ from the per-night noise-wave solution \citep{Roque2021}, are similarly attached. These metadata are not used as inputs to the stage-1 or stage-2 likelihoods; they are joined to the observation frame after the fits for post-hoc correlation analysis and validation.

% ----------------------------------------------------------------------------
\subsection{Stage-1 implementation}
\label{subsec:stage1_implementation}
% ----------------------------------------------------------------------------

Each of the $T = 4655$ observations, spanning one year of REACH operation, is fit independently using the one-order chromaticity model of equation~(\ref{eq:one_order}) with the per-channel anomaly mixture of equation~(\ref{eq:channel_marginal}). The sampled parameters for each observation are the foreground amplitude $A_t \sim \mathcal{U}[0.5, 2]$, the spectral index $\beta_t \sim \mathcal{U}[0, 4]$, the log-noise scale $\log_{10}\sigma_t \sim \mathcal{U}[-1.5, 1.5]$, and the log-anomaly probability $\log_{10}P_{a,1} \sim \mathcal{U}[-6, -0.5]$, with the anomaly support width $\Delta_{a,1}$ set to approximately the maximum observed antenna temperature.

Each fit is performed using \blackjax\ NSS \citep{Yallup2026nss,Cabezas2024} within the \reach\ GPU pipeline \citep{Tutt2026}. The number of live points is scaled as $n_{\rm live} = 20\,n_{\rm dims}$, the number of inner slice-sampling steps as $n_{\rm inner} = 6\,n_{\rm dims}$, and the deletion fraction as $0.5\,n_{\rm live}$. The stopping criterion terminates sampling when the estimated remaining log-evidence contribution from the live points satisfies $\log \mathcal{Z}_{\rm live} - \log \mathcal{Z} < -3$.

For each completed stage-1 fit, the following quantities are saved: $K = 256$ posterior draws of $\tref$ computed via equation~(\ref{eq:tref_draw}), summary statistics ($\tref^{\rm mean}$, $\tref^{\rm std}$), per-channel anomaly probabilities via equation~(\ref{eq:channel_responsibility}), and the mean anomaly probability averaged across frequency channels.

% ----------------------------------------------------------------------------
\subsection{Stage-2 implementation}
\label{subsec:stage2_implementation}
% ----------------------------------------------------------------------------

The stage-2 Fourier model of equation~(\ref{eq:fourier}) is fit to the full ensemble of $\tref$ posteriors using a second \blackjax\ NSS run. The sampled parameters are the Fourier coefficients $\bm{c} = \{c_0, a_h, b_h\}$, the log-intrinsic scatter $\log_{10}\sigma_{\rm int}$, and the log-anomaly probability $\log_{10}P_{a,2}$. The prior bounds are derived from the stage-1 posterior summaries: $c_0 \sim \mathcal{U}[\tref^{\rm min}, \tref^{\rm max}]$ and $a_h, b_h \sim \mathcal{U}[-c_{\rm max}, c_{\rm max}]$, where $\tref^{\rm min}$ and $c_{\rm max}$ are computed from the propagated $\tref$ draws and $\tref^{\rm max}$ is the configured upper bound. For the run using the evidence-preferred $H=3$ model identified in Section~\ref{subsec:model_selection}, $\tref^{\rm min} = 226.77$\,K and $c_{\rm max} = 5829.92$\,K.

For each harmonic order $H$, the model has $n_{\rm dims} = 2H + 3$ parameters, and the nested sampling configuration scales as $n_{\rm live} = 10\,n_{\rm dims}$ live points with $n_{\rm inner} = 4\,n_{\rm dims}$ inner slice-sampling steps. For this evidence-preferred $H = 3$ model, this gives $n_{\rm live} = 90$ and $n_{\rm inner} = 36$.

The stage-2 posterior anomaly probability for each observation is computed by averaging over the posterior samples. For each posterior sample $s$ with parameters $(\bm{c}^{(s)}, P_{a,2}^{(s)}, \sigma_{\rm int}^{(s)})$, the draw-averaged anomaly responsibility for observation $t$ is
\begin{multline}
\label{eq:stage2_anomaly_prob}
\hat{P}_{a,2}(t) = \frac{1}{N_{\rm post}} \sum_{s=1}^{N_{\rm post}} \frac{1}{K}\sum_{k=1}^{K} \\
\frac{P_{a,2}^{(s)}\,\Delta_{a,2}^{-1}}{(1-P_{a,2}^{(s)})\,\mathcal{N}(\treftk \mid \mu_t^{(s)},\, \sigma_{t}^{(s)\,2}) + P_{a,2}^{(s)}\,\Delta_{a,2}^{-1}},
\end{multline}
where $\sigma_t^{(s)\,2} = \sigma_{t,\mathrm{ker}}^2 + (\sigma_{\rm int}^{(s)})^2$ and $\mu_t^{(s)}$ is the Fourier curve evaluated at the LST of observation $t$ under sample $s$.

% ----------------------------------------------------------------------------
\subsection{Fourier-order selection}
\label{subsec:fourier_selection}
% ----------------------------------------------------------------------------

The number of Fourier harmonics $H$ is not chosen a priori. Instead, we perform independent stage-2 nested sampling runs for $H = 1, 2, \dots, 6$ and compare the resulting Bayesian evidences $\mathcal{Z}_H$. The model with the highest evidence is selected as the preferred description of $\tref(\mathrm{LST})$. Because nested sampling delivers both the posterior and the evidence as standard outputs \citep{Skilling2006}, this model-selection step requires no additional computation beyond running the suite of fits. The evidence uncertainties are estimated from 128 realisations of the nested-sampling compression weights \citep{Skilling2006}, yielding a mean and standard deviation for $\log\mathcal{Z}_H$.

% ----------------------------------------------------------------------------
\subsection{Computational infrastructure}
\label{subsec:computational}
% ----------------------------------------------------------------------------

The entire pipeline is implemented in Python using \jax\ \citep{Bradbury2018} for array computation and automatic differentiation, and \blackjax\ \citep{Cabezas2024} for nested sampling. All stage-1 and stage-2 fits are executed on GPU, exploiting the vectorised nested slice sampling algorithm of \citet{Yallup2026nss} to parallelise the live-point evolution across GPU threads. Post-processing and visualisation use \textsc{anesthetic} \citep{Handley2019anesthetic} for posterior resampling and \fgivenx\ \citep{Handley2018fgivenx} for functional posterior plotting. The pipeline is designed to run without human intervention from raw observation data to final anomaly-probability assignments and figure generation.

% ============================================================================
\section{Results}
\label{sec:results}
% ============================================================================

We present the results of applying the fully automatic curation pipeline to the \reach\ first-year dataset. The through-line of this section is that probabilistic flagging at scale is useful only if it both protects inference from contaminated observations and produces diagnostics that help researchers identify why the data are anomalous. We therefore first establish the automatically selected population model, then show the large-scale anomaly trends in frequency and time, connect those flags to known instrumental and environmental causes, and finally summarise what this enables for \reach. All subsequent results in this section use the evidence-preferred $H=3$ stage-2 model.

% ----------------------------------------------------------------------------
\subsection{Fourier-order model selection}
\label{subsec:model_selection}
% ----------------------------------------------------------------------------

\begin{figure}
\centering
\includegraphics[width=\columnwidth]{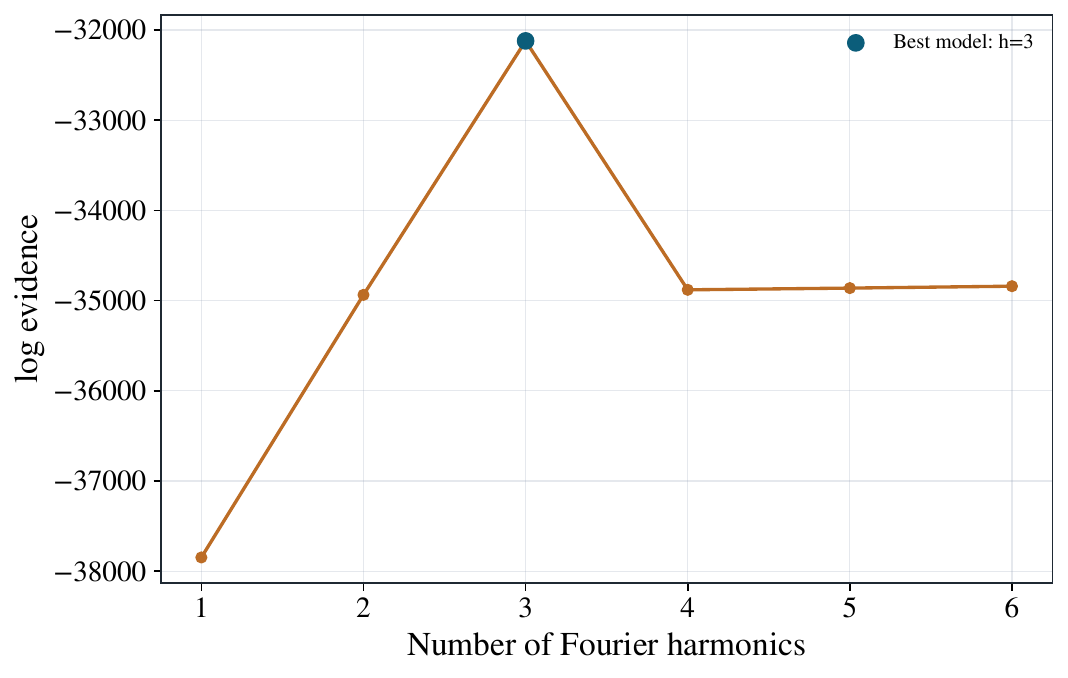}
\caption{Bayesian evidence versus number of Fourier harmonics. The $H = 3$ model is decisively preferred; the sharp peak confirms automatic model complexity selection via the Occam penalty.}
\label{fig:evidence}
\end{figure}

Fig.~\ref{fig:evidence} shows the Bayesian evidence as a function of Fourier harmonic order. The evidence peaks sharply at $H = 3$, with all other models decisively disfavoured due to the Bayesian Occam penalty. This confirms that three harmonics provide sufficient flexibility to capture the dominant LST structure, while additional complexity is not justified by the data. The selection is made automatically by the evidence without human intervention.

With the LST model complexity fixed by the evidence, the anomaly probabilities can be interpreted against a population trend chosen by the data rather than by manual tuning.

% ----------------------------------------------------------------------------
\subsection{General trends in frequency and time}
\label{subsec:general_trends}
% ----------------------------------------------------------------------------

The first role of the anomaly scores is to replace manual visual triage with a reproducible map of where the data depart from the inferred clean population. This is the scale at which analyst time is saved: the pipeline reduces thousands of spectra to probabilistic summaries across frequency, LST, and calendar time.

\subsubsection{Antenna temperature and anomaly waterfall}
\label{subsec:waterfall}

\begin{figure*}
\centering
\includegraphics[width=\textwidth]{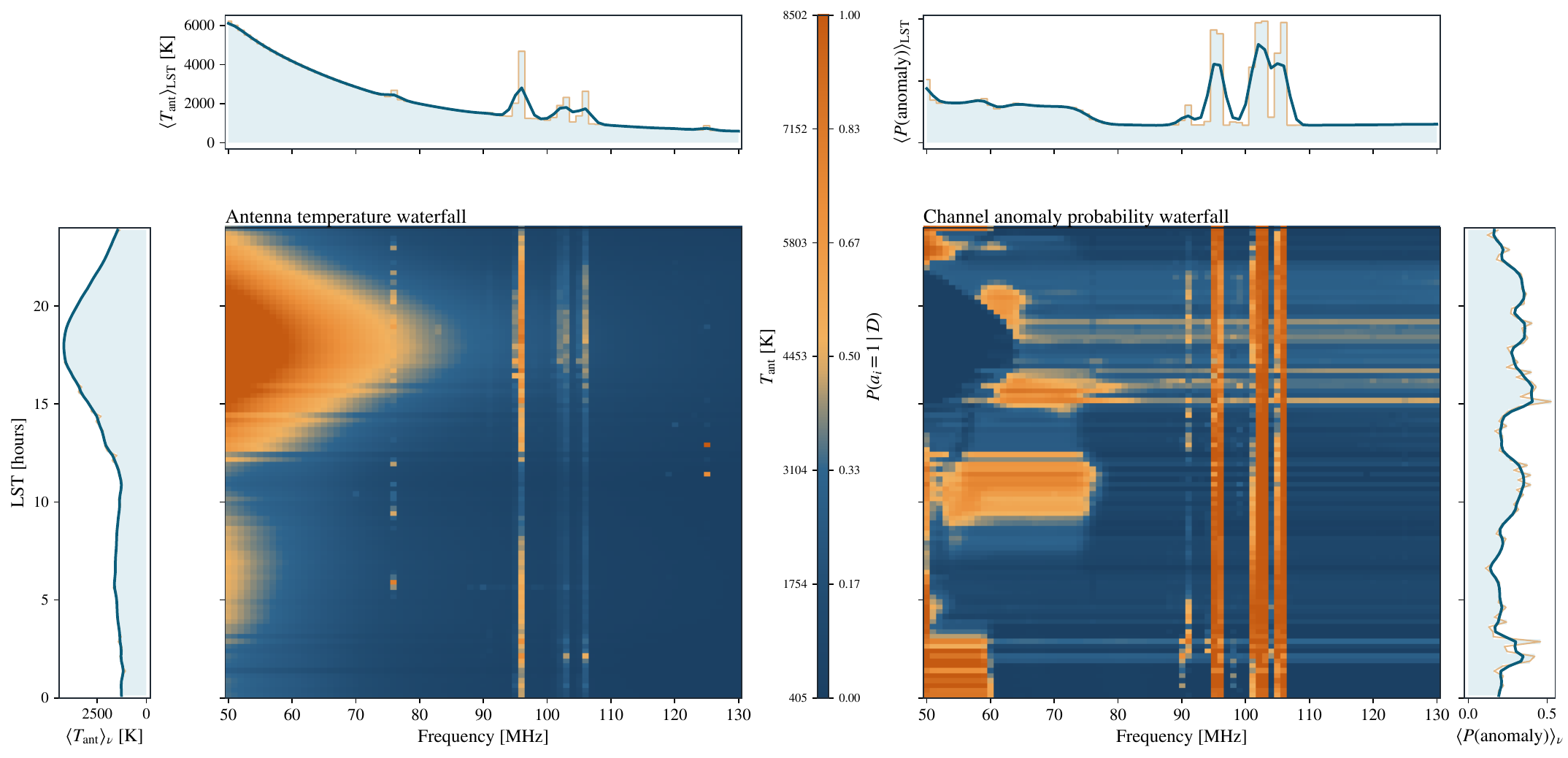}
\caption{Antenna temperature (left) and channel-level anomaly probability (right) as functions of frequency and LST. The antenna-temperature waterfall shows the expected synchrotron emission modulated by beam chromaticity; the anomaly waterfall reveals persistent contaminated channels ($95$--$106$\,MHz) and an LST-dependent contamination gradient.}
\label{fig:waterfall}
\end{figure*}

Fig.~\ref{fig:waterfall} provides the broadest view of the dataset. The left panel shows the antenna-temperature waterfall as a function of frequency and LST, dominated by the bright low-frequency synchrotron emission that traces the Galactic plane as it transits overhead. The right panel shows the corresponding channel-level anomaly-probability waterfall. The two maps reveal qualitatively different structure: where the antenna temperature varies smoothly with the sky, the anomaly waterfall is dominated by persistent vertical stripes at a small number of contaminated frequency channels, most prominently in the $95$--$106$\,MHz range. The anomaly burden also increases with LST through the middle and later parts of the sidereal day.

\subsubsection{The \texorpdfstring{$\tref(\mathrm{LST})$}{Tref(LST)} posterior and observation-level anomaly scores}
\label{subsec:tref_posterior}

\begin{figure*}
\centering
\includegraphics[width=\textwidth]{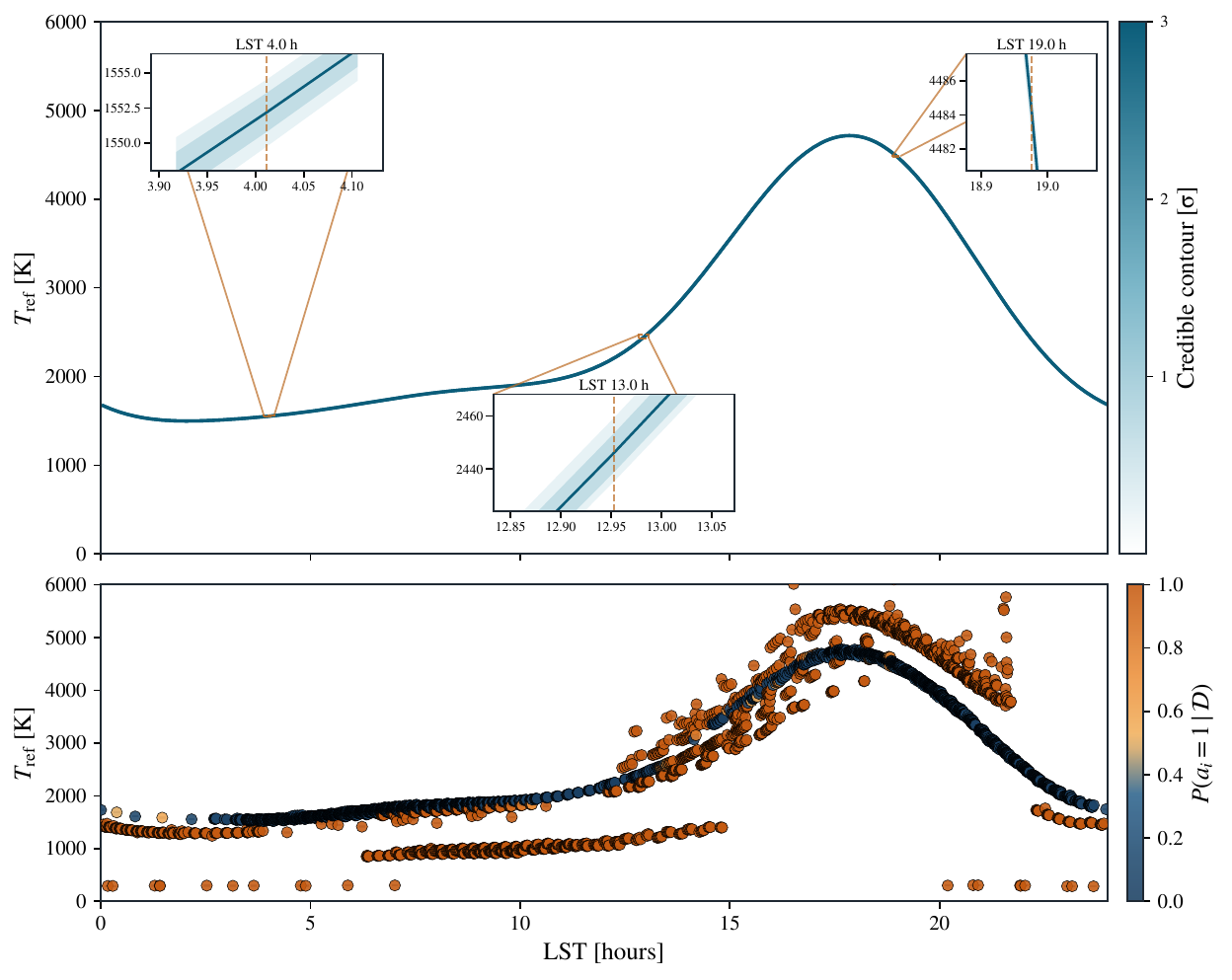}
\caption{\textit{Top:} posterior distribution of the smooth $\tref(\mathrm{LST})$ Fourier curve shown as \fgivenx\ credible contours. \textit{Bottom:} per-observation $\tref$ estimates coloured by anomaly probability. The latent curve is tightly constrained despite the far larger scatter in individual estimates, and high-anomaly observations are concentrated among the outliers.}
\label{fig:tref_posterior}
\end{figure*}

Fig.~\ref{fig:tref_posterior} shows the stage-2 posterior over the smooth latent $\tref(\mathrm{LST})$ curve together with per-observation summaries. The top panel displays \fgivenx\ credible contours ($1\sigma$, $2\sigma$, $3\sigma$), providing properly informed uncertainty envelopes rather than simple point-estimate error bars. The smooth curve traces the expected diurnal variation as the Galactic plane transits through the beam, and the posterior is very tightly constrained, with an inferred intrinsic scatter of only $\sigma_{\rm int} = 22.36$\,K.

The bottom panel shows per-observation $\tref$ estimates coloured by anomaly probability. A substantial fraction of observations are flagged as anomalous and scattered far from the smooth curve, while the majority cluster tightly around it. Notably, there are also borderline observations that to the eye appear consistent with the curve but are assigned elevated anomaly probabilities by the model. These represent cases where the Bayesian framework identifies subtle inconsistencies that might be missed by visual inspection and would otherwise be available to bias downstream inference.

\subsubsection{Frequency-channel anomaly structure}
\label{subsec:channel_anomaly}

\begin{figure*}
\centering
\includegraphics[width=\textwidth]{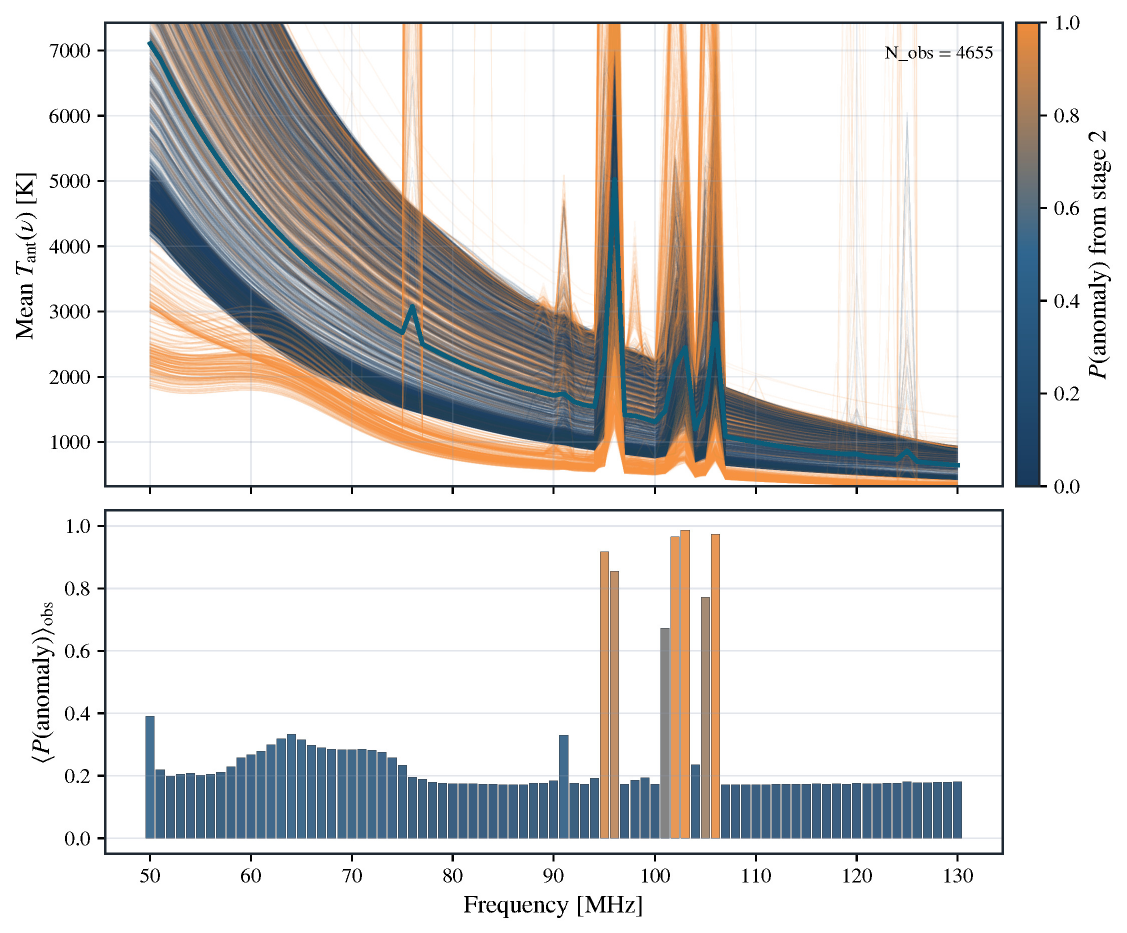}
\caption{\textit{Top:} mean antenna-temperature spectrum with individual observations coloured by anomaly probability. \textit{Bottom:} mean anomaly probability per frequency channel. Persistent contamination is concentrated at $95$--$106$\,MHz (FM band), while transient RFI is visible as intermittently flagged channels.}
\label{fig:channel_anomaly}
\end{figure*}

Fig.~\ref{fig:channel_anomaly} examines contamination in the spectral domain. The top panel overlays individual observation spectra, coloured by anomaly probability, on the mean antenna-temperature spectrum. The bottom panel shows the mean anomaly probability per frequency channel.

The pipeline identifies the $95$--$106$\,MHz region as the dominant source of persistent narrow-band contamination, consistent with known RFI from the FM broadcast band at the Karoo site. The band-edge channel at $50$\,MHz is also elevated. Critically, the top panel reveals that some channels exhibit high anomaly probability in only a subset of observations, indicating that the pipeline detects transient as well as persistent RFI. This is an important capability: transient interference that contaminates a channel intermittently is harder to identify than persistent features, yet the Bayesian anomaly framework flags both without requiring separate detection strategies.

\subsubsection{Temporal anomaly structure}
\label{subsec:temporal}

\begin{figure*}
\centering
\includegraphics[width=\textwidth]{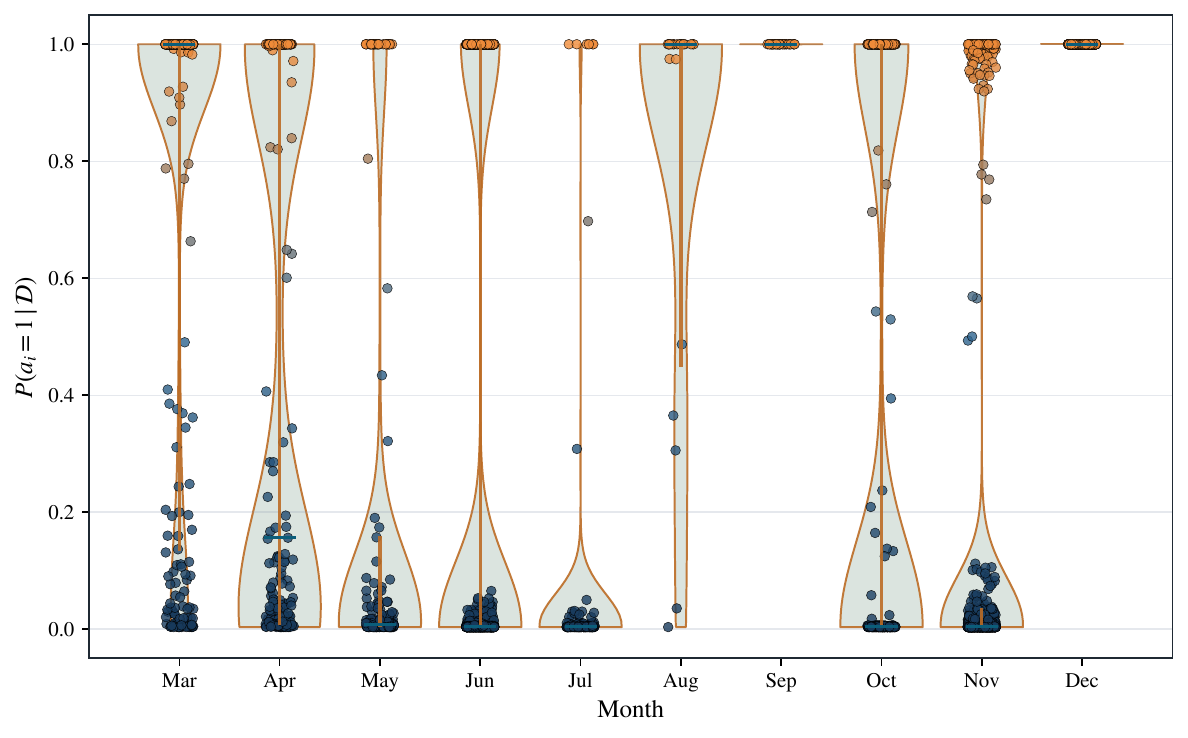}
\caption{Stage-2 anomaly probability versus calendar month. March, August, September, and December are dominated by high anomaly scores; July is the cleanest month. Even in predominantly clean months, individual contaminated observations are identified.}
\label{fig:month}
\end{figure*}

\begin{figure}
\centering
\includegraphics[width=\columnwidth]{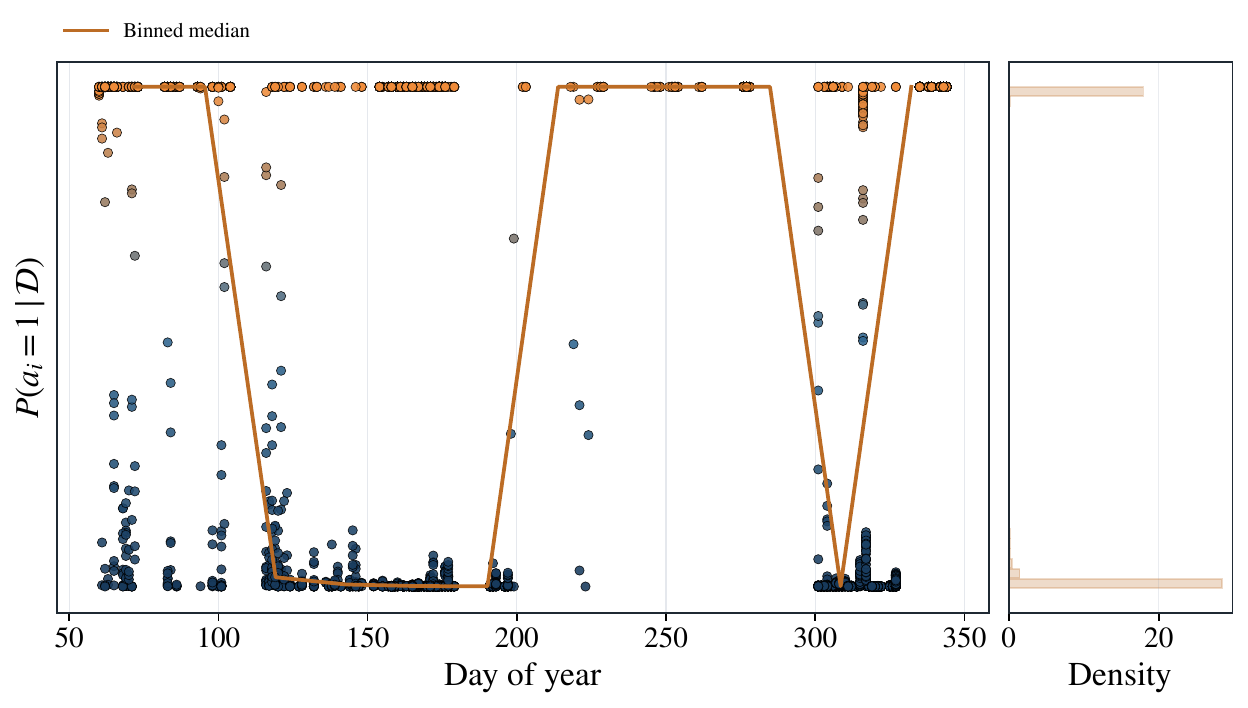}
\caption{Stage-2 anomaly probability versus day of year. The contamination pattern is episodic, with several strongly affected windows separated by a mid-year clean trough.}
\label{fig:doy}
\end{figure}

Figs~\ref{fig:month} and~\ref{fig:doy} show the distribution of anomaly probability as a function of calendar month and day of year respectively. The contamination pattern is strongly episodic rather than smoothly seasonal. This behaviour is expected for these data, because much of the contamination arises from instrumental faults that were subsequently repaired; a smooth seasonal progression of anomalies is therefore not expected. March, August, September, and December are dominated by high anomaly scores, while July is the cleanest month. June and October contain mixtures of clean and contaminated populations: their medians are low but they exhibit substantial high-probability tails. Critically, even within months that are predominantly clean, individual observations are flagged as anomalous, demonstrating that the pipeline identifies localised contamination events that a simple month-level rejection strategy would miss.

The finer day-of-year resolution (Fig.~\ref{fig:doy}) confirms this episodic structure, with several strongly contaminated windows separated by a clean mid-year trough. These temporal flags are the bridge from curation to diagnosis: once anomalous windows are identified automatically, they can be cross-matched against independent records of instrumental state and observing conditions.

% ----------------------------------------------------------------------------
\subsection{Correlation with known causes of systematics}
\label{subsec:known_causes}
% ----------------------------------------------------------------------------

The second role of large-scale flagging is to turn anomaly probabilities into a diagnostic tool. Because the metadata considered below are not used in the likelihood, correlations with them provide blind validation and help identify the physical causes of contamination. This diagnostic step is a post-hoc analysis performed by researchers with domain knowledge: the pipeline automatically assigns anomaly probabilities and downweights contaminated data in the inference, while humans subsequently compare those scores with metadata such as calibration state and weather.

\subsubsection{Anomaly probability versus noise-source temperature}
\label{subsec:tns}

\begin{figure*}
\centering
\includegraphics[width=\textwidth]{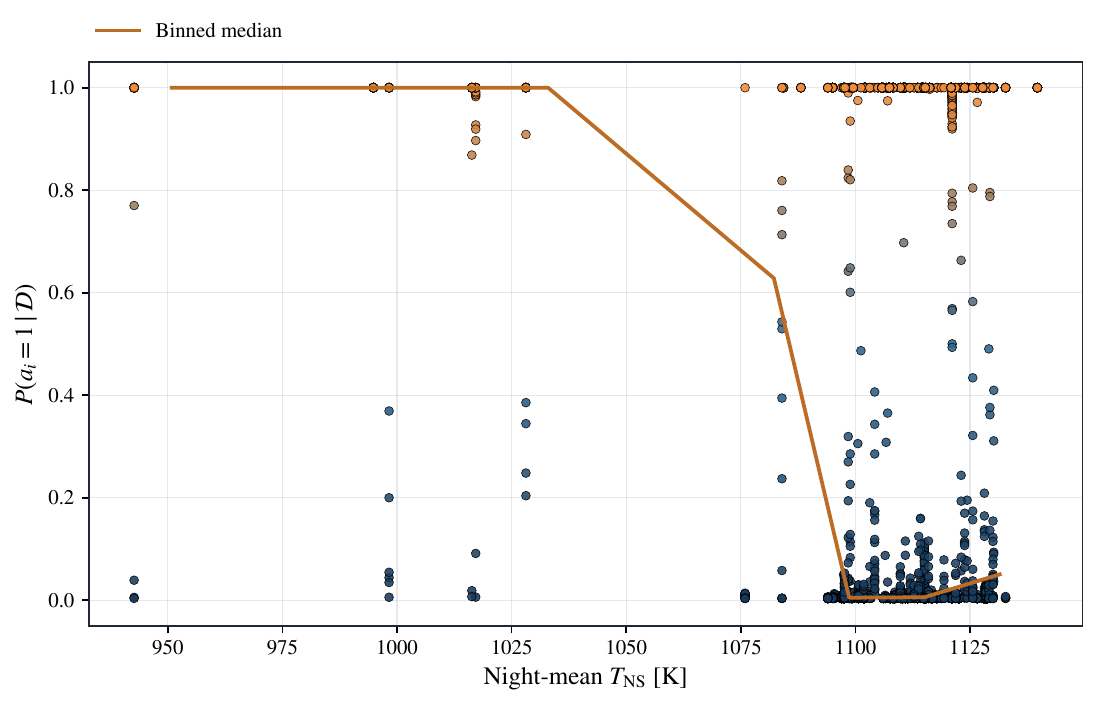}
\caption{Stage-2 anomaly probability versus night-mean noise-source temperature $\tns$. The pipeline has no knowledge of $\tns$ during inference; the correlation is identified in a post-hoc comparison between automatic anomaly scores and calibration metadata.}
\label{fig:tns}
\end{figure*}

Fig.~\ref{fig:tns} tests whether the pipeline's anomaly scores correlate with the state of the calibration hardware. The noise-source diode is an internal component whose temperature $\tns$ is used to map raw digitiser counts to antenna temperature on the sky \citep{Roque2025,Roque2021}. The pipeline has no knowledge of $\tns$ during inference; the comparison below is a human post-hoc cross-match between automatic anomaly scores and calibration metadata.

The relationship is striking. At the approximate correct operating value of $\tns$, observations are scored as clean; at values that deviate from this, data quality degrades sharply and the pipeline assigns high anomaly probabilities. Not all observations close to the correct $\tns$ value are clean, however: many are still flagged as anomalous, reflecting other sources of anomalous data beyond noise-source miscalibration. Low-to-moderate $\tns$ nights are almost entirely flagged as anomalous. An incorrect $\tns$ propagates directly into the calibrated spectrum and shifts $\tref$ away from the smooth population trend, so this correlation has a clear physical explanation. The curation itself does not depend on recognising this cause: the contaminated observations are already downweighted inside the likelihood, and the $\tns$ comparison provides subsequent human validation and diagnosis.

\subsubsection{Anomaly probability versus precipitation}
\label{subsec:precipitation}

\begin{figure}
\centering
\includegraphics[width=\columnwidth]{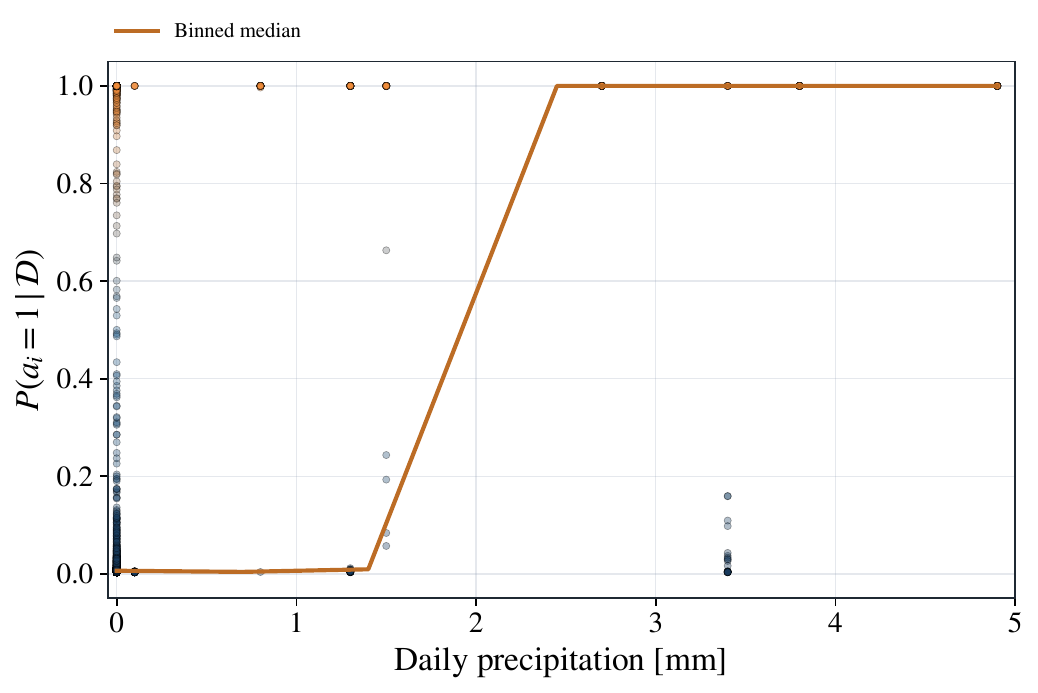}
\caption{Binned-median anomaly probability versus daily precipitation over the well-sampled $0$--$5\,\mathrm{mm}$ range. Rainy days are associated with elevated anomaly probability, despite precipitation not being supplied to the pipeline.}
\label{fig:precipitation}
\end{figure}

Fig.~\ref{fig:precipitation} examines the relationship between daily precipitation at the Karoo site and anomaly probability. As with $\tns$, precipitation is not supplied to the pipeline at any stage; the association is identified by a post-hoc comparison of automatic anomaly scores with weather metadata. Rain is expected to degrade data quality through changes in antenna impedance and atmospheric opacity \citep{Pattison2025env}, and rainy days are associated with substantially elevated anomaly scores. The relationship is non-monotonic, with moderate precipitation most strongly associated with high anomaly probability. The plotted summary is the binned median anomaly probability, shown only over the well-sampled $0$--$5\,\mathrm{mm}$ precipitation range. Regardless of whether this weather correlation is inspected, the affected observations have already been probabilistically downweighted by the inference.

Together, the $\tns$ and precipitation correlations show that the anomaly scores do more than rank data quality. They provide a route from statistical outliers to instrumental or environmental causes, when analysts compare the automatic flags with external metadata, allowing systematics to be investigated and corrected rather than merely discarded. This diagnostic interpretation is human-led, but the curation action, assigning anomaly probabilities and marginalising over contaminated observations, is automatic.

% ----------------------------------------------------------------------------
\subsection{Why flagging at scale during inference is important for REACH}
\label{subsec:why_scale_matters}
% ----------------------------------------------------------------------------

For \reach, the practical value of this pipeline is that it makes data selection a reproducible inference product rather than a manual preprocessing decision. A first-year dataset of 4655 observations is still small enough for extensive human inspection, but the relevant task is not simply to notice obvious bad nights. The analysis must prevent anomalous observations from biasing the inferred foreground and global-signal model, while retaining genuinely informative data wherever possible. Probabilistic anomaly scores allow this trade-off to be made inside the likelihood, with uncertainty carried forward instead of collapsed into a hard threshold.

This also changes how researchers interact with the experiment. Rather than inspecting thousands of spectra by hand, analysts can identify the observing periods, frequency channels, and hardware states most responsible for degraded data quality, then compare those flags with independent environmental and instrumental records. The results above show all three uses simultaneously: the pipeline finds large-scale contamination structure, protects the smooth $\tref(\mathrm{LST})$ inference from outliers, and recovers correlations with $\tns$ and precipitation that were not supplied to the model. The same capabilities are essential for larger instruments, where manual curation is infeasible and the number of possible systematic causes is much larger.

% ============================================================================
\section{Discussion}
\label{sec:discussion}
% ============================================================================

The results demonstrate that Bayesian anomaly detection can move data curation from external flagging into the inference itself. Applied at the scale of a full observing season, the method recovers known contamination structure in the \reach\ first-year dataset without human intervention, while marginalising over the possibility that individual observations or channels are contaminated.

% ----------------------------------------------------------------------------
\subsection{Scalable automation of data curation}
\label{subsec:scalable_automation}
% ----------------------------------------------------------------------------

The pipeline automates a process that has traditionally required substantial human effort, but its main advantage is statistical rather than only operational. Identifying contaminated months, flagging narrow-band RFI, and assessing weather-related degradation are tasks that experienced analysts already perform, but they do so by inspecting diagnostic plots, applying ad hoc thresholds, and exercising subjective judgement. By embedding curation in the likelihood, the same decisions become posterior quantities that can protect parameter inference from contaminated data without converting uncertainty into a hard pre-analysis mask. For datasets of the size analysed here, manual curation is laborious but feasible. For the next generation of radio experiments it will not be.

The \ska\ will comprise hundreds of thousands of individual antenna elements and produce raw data volumes exceeding an exabyte per day \citep{Dewdney2009}. \lofar\ already manages petabyte-scale archives \citep{vanHaarlem2013,Shimwell2019}, and \hera\ records terabytes per night \citep{DeBoer2017}. At these scales, the number of potential contamination pathways grows combinatorially with the complexity of the instrument: interactions between receiver components, correlator artefacts, site-specific environmental effects, and time-dependent calibration drifts can all produce subtle data-quality degradation that is difficult for simple algorithms or human inspectors to identify. Bayesian anomaly detection provides a route to automated curation that scales naturally with data volume and instrument complexity, because the contamination model is embedded in the likelihood rather than imposed as an external preprocessing step. The science gain is correspondingly direct: cleaner and less biased inference can be pursued without manual thresholds that become impossible to audit at survey scale.

% ----------------------------------------------------------------------------
\subsection{Discovery of complex contamination relationships}
\label{subsec:discovery}
% ----------------------------------------------------------------------------

Beyond automating existing curation practice, the pipeline creates anomaly products that help researchers discover contamination relationships that would be difficult to uncover through manual inspection alone. The correlation between anomaly probability and noise-source temperature (Fig.~\ref{fig:tns}) is a clear example: $\tns$ is not provided to the likelihood, yet the automatic anomaly scores reveal a strong relationship when human analysts compare them with calibration metadata. This goes beyond what a simple flagging algorithm could achieve, because the relationship between $\tns$ and data quality is mediated through the calibration chain in a way that is not obvious from the raw data alone.

Similarly, the precipitation relationship (Fig.~\ref{fig:precipitation}) and the detection of transient RFI alongside persistent interference (Fig.~\ref{fig:channel_anomaly}) demonstrate that the Bayesian framework captures contamination structure across multiple domains simultaneously. Identifying the physical causes of these structures still requires domain expertise and post-hoc comparison with external records, but the removal or downweighting of contaminated data does not: it has already happened probabilistically inside the inference. For instruments with hundreds of thousands of components, such as the \ska, anomaly scores that can be compared against many streams of hardware and environmental metadata could prove at least as valuable as the automated curation itself.

The Fourier-order selection (Fig.~\ref{fig:evidence}) illustrates a further advantage: model complexity is set by the Bayesian evidence rather than by human judgement \citep{MacKay2003,Jeffreys1961}, eliminating a class of analyst degrees of freedom that introduce subjectivity into conventional pipelines.

% ----------------------------------------------------------------------------
\subsection{The two-stage approximation}
\label{subsec:two_stage}
% ----------------------------------------------------------------------------

The two-stage design adopted here is a trade-off between statistical fidelity and computational tractability. A fully joint model over all observations simultaneously would retain more information but would introduce thousands of nuisance parameters. The two-stage approximation compresses each observation to a posterior over $\tref$, discarding the full channel-level anomaly pattern but retaining its dominant effect on the compressed summary. The tight posterior constraints on the latent curve ($\sigma_{\rm int} = 22.36$\,K) confirm that this approximation is effective in practice.

This is a minor limitation. A natural extension would incorporate models for time-variant systematics \citep{Kirkham2024} to capture instrumental drifts on timescales shorter than the LST bin width, and Laplace-marginalised approaches \citep{Kirkham2025} could retain more stage-1 structure without requiring the full joint sampling.

% ============================================================================
\section{Conclusion}
\label{sec:conclusion}
% ============================================================================

We have presented a fully automatic data curation method based on Bayesian anomaly detection \citep{Leeney2023,Anstey2024rfi} and demonstrated it on 4655 observations from the \reach\ first-year dataset. The pipeline requires no flagging thresholds, no manual inspection, and no subjective quality decisions. Applied to real data, it identifies contamination across frequency, time, and local sidereal time, recovering known contaminants including narrow-band RFI in the FM broadcast band, miscalibrated noise-source temperatures, and precipitation-driven systematics, all without prior knowledge of these quantities. The contamination structure is episodic rather than smoothly seasonal, and the pipeline captures both persistent and transient effects.

Beyond reproducing what experienced analysts would flag manually, the method discovers complex contamination relationships that would be difficult to identify by inspection, such as the correlation between anomaly probability and the internal noise-source temperature. This capability becomes increasingly important as instruments grow in complexity. The methodology is domain-general: it has already been applied to Type Ia supernova cosmology \citep{Leeney2025snia}, and the principles extend naturally to any observational programme in which automated, uncertainty-aware data curation is required.

As radio astronomy enters an era in which data volumes preclude manual curation \citep{Dewdney2009,vanHaarlem2013,DeBoer2017}, scalable Bayesian data curation will become not merely convenient but necessary. This work demonstrates that it is achievable in practice.

% ============================================================================
\section*{Acknowledgements}
% ============================================================================

S.A.K. Leeney led the research, development, and preparation of the manuscript, with E. de Lera Acedo, W.J. Handley and H.T.J. Bevins assisting. The remaining authors, from G. Allen to J. Zhu, contributed to the REACH telescope, and thus are on the REACH builders list. They also contributed to the manuscript's preparation and review. These authors are credited in alphabetical order.

% TODO: funding/grant acknowledgements still to be added (no funding statement existed in the radiometer-calibration ML paper to copy).

% ============================================================================
% BIBLIOGRAPHY
% ============================================================================

\bibliographystyle{mnras}

\label{lastpage}
\end{document}